\documentclass[12pt]{article}

\usepackage{latexsym}
\usepackage{amsmath,amssymb}

\usepackage[bookmarks=false]{hyperref}

\newcommand{\be}{\begin{equation}}
\newcommand{\ee}{\end{equation}}

\newcommand{\1}{{\bf1}}
\newcommand{\2}{{\bf2}}
\newcommand{\3}{{\bf3}}
\newcommand{\4}{{\bf4}}

\newcommand{\ald}{{\dot\alpha}}
\newcommand{\Dbar}{\bar D{}}

\def\tr{{\rm tr}\,}

\def\cN{{\cal N}}
\def\dfrac{\displaystyle \frac }

\newcommand{\bea}{\begin{eqnarray}}
\newcommand{\eea}{\end{eqnarray}}
\newcommand{\ba}{\begin{array}}
\newcommand{\ea}{\end{array}}
\newcommand{\nn}{\nonumber}

\newcommand{\p}{\partial}

\newcommand{\tabar}{\bar\ta{}}

\newcommand{\al}{\alpha}

\newcommand{\si}{\sigma}
\newcommand{\ta}{\theta}

\begin{document}

\begin{titlepage}

\vspace{1cm}

\begin{center}
{\Large\bf Bi-harmonic superspace \\[3mm]
for $\cN=4$\, $d=4$ super Yang-Mills} \vspace{1.5cm}

{\large\bf Dmitry~V.~Belyaev\,{}$^\dag$ {\it and}\hspace{5pt}
 Igor~B.~Samsonov$\,{}^{\star}$\footnote{On leave from
Tomsk Polytechnic University, 634050 Tomsk, Russia}
\\[8pt]
\it\small $^\dag$ Institute for Fundamental Theory, Department of Physics,
\\ University of Florida, Gainesville, FL 32611, USA\\
{\tt email:\ belyaev@phys.ufl.edu}\\[8pt]
$^\star$INFN, Sezione di Padova, via F. Marzolo 8, 35131 Padova, Italy\\
{\tt email:\ samsonov@mph.phtd.tpu.ru}}
\end{center}
\vspace{0.5cm}

\begin{abstract}
We develop $\cN=4$\, $d=4$ bi-harmonic superspace and use it to
derive a novel form for the low-energy effective action in $\cN=4$
super Yang-Mills theory. We solve the $\cN=4$ supergauge
constraints in this superspace in terms of analytic superfields.
Using these superfields, we construct a simple functional that
respects $\cN=4$ supersymmetry and scale invariance. In
components, it reproduces all on-shell terms in the four-derivative part of
the $\cN=4$ SYM effective action; in particular, the
$F^4/X^4$ and Wess-Zumino terms. The latter comes out in a novel
SO(3)$\times $SO(3)-invariant form.
\end{abstract}

\end{titlepage}

\tableofcontents

\numberwithin{equation}{section}

\section{Introduction}

In spite of many attempts, the problem of formulating  the
classical action of the $\cN=4$ super Yang-Mills (SYM) theory in
$\cN=4$ superspace remains unsolved. In our recent paper
\cite{BS}, however, we demonstrated that the $\cN=4$ USp(4)
harmonic superspace can be naturally used to describe the
low-energy {\it effective} action of this model. We considered
there leading terms in the derivative expansion of the $\cN=4$ SYM
low-energy effective action on the Coulomb branch, which are given
by the so-called `$F^4/X^4$' term \cite{DS,S16}
and the Wess-Zumino (WZ) term \cite{TZ,Intriligator:2000eq}, and showed
that they originate from a simple functional in the $\cN=4$
harmonic superspace with USp(4) harmonic variables. The low-energy
effective action in this superspace has a remarkably simple form
owing to scale invariance and explicit $\cN=4$ supersymmetry
\cite{BLS}. In this paper, we will describe another superspace
where the form of the effective action turns out to be equally
simple.

The WZ term in the $\cN=4$ SYM effective action
\cite{TZ,Intriligator:2000eq} can be written in a manifestly
SO(6)$\sim$SU(4)-invariant form at the price of sacrificing
locality: by writing it as an integral over a five-dimensional
manifold that has the four-dimensional Minkowski space as its
boundary. In a local four-dimensional form of the WZ term, only a
subgroup of SU(4) can be manifest \cite{Claus,Braaten}. In
\cite{BS}, we presented three different forms of the WZ term in
the four-dimensional Minkowski space, which are manifestly
invariant under SO(5), SO(4)$\times$SO(2), and SO(3)$\times$SO(3),
respectively. These are nothing but the three maximal
non-anomalous subgroups of the SU(4) R-symmetry group. We argued
there that the most elegant description of the $\cN=4$ low-energy
effective action must be in those superspaces which make these
subgroups manifest.

In \cite{BS}, we matched two of the non-anomalous subgroups, SO(5)
and SO(4)$\times$SO(2), with two known superspace descriptions of
the $\cN=4$ SYM effective action. The former corresponds to the
$\cN=4$ USp(4) harmonic superspace \cite{BLS}, whereas the latter
to the standard $\cN=2$ SU(2) harmonic superspace
\cite{GIKOS,book}.~\footnote{ The $\cN=4$ SYM effective action in
the $\cN=2$ harmonic superspace was first constructed in
\cite{BuIv}, and later rederived through direct perturbative
computations in \cite{BIP,BBP,BP1,BP2}.} However, no superspace
was known which would make the SO(3)$\times$SO(3) subgroup
manifest. In this paper, we will fill the gap by introducing
$\cN=4$\, $d=4$ bi-harmonic superspace with explicitly realized
SU(2)$\times$SU(2)$\sim$SO(3)$\times$SO(3) subgroup of the
SU(4)$\sim$SO(6) R-symmetry group.~\footnote{We note that similar
bi-harmonic superspaces have been used in
\cite{IS1,Ivanov:1995jb,Ivanov:1995yp,BI,IS2} to describe $d=2$ supersymmetric
sigma models and in \cite{BIS,IN} to discuss $d=1$ supersymmetric
mechanics with extended supersymmetry.} We will demonstrate that the
description of the $\cN=4$ SYM low-energy effective action in this
superspace is as elegant as in the USp(4) harmonic superspace
\cite{BS,BLS}.

The six real scalars $X^M$, $M=1,\ldots,6$ in the $\cN=4$ SYM theory
transform as a vector of the SO(6) R-symmetry group. So do the
normalized scalars $Y^M=X^M/|X|$, which lie on the unit sphere.
Splitting them into two triplets, $Y^M=(Y^A,Y^{A'})$, $A,A'=1,2,3$, we
find that $Y^A$ and $Y^{A'}$ transform as vectors under different
SO(3) subgroups of the SO(6) group.  In \cite{BS}, we showed that
the WZ term of the $\cN=4$ SYM effective action can then be written as
\be
-\frac1{16\pi^2}\varepsilon^{mnpq}
\int d^4x\, g(y) (\varepsilon_{ABC}Y^A\partial_m Y^B \partial_n
Y^C)
(\varepsilon_{{A'}{B'}{C'}} Y^{A'} \partial_p Y^{B'} \partial_q Y^{C'})\,,
\label{1}
\ee
where
\be
g(y)=\frac{y^4-1}{y^2}+\frac{(y^2+1)^3}{y^3}\arctan y\,,\qquad
y^2=\frac{Y^A Y^A}{Y^{A'}Y^{A'}}\,.
\label{2}
\ee
In this form, the WZ term is explicitly invariant under the
SO(3)$\times$SO(3) subgroup of SO(6), whereas the rest of the SO(6)
invariance is implicit.  As we will see, the WZ term (\ref{1}) follows
naturally from a functional in the $\cN=4$\, $d=4$ bi-harmonic superspace.

This paper is organized as follows. In Section 2, we introduce
SU(2)$\times$SU(2) harmonic variables in the standard $\cN=4$\,
$d=4$ superspace and classify analytic subspaces in the resulting
bi-harmonic superspace. In Section 3, we define analytic (short)
superfields in this superspace which solve a part of the $\cN=4$
supergauge constraints. Section 4 is devoted to the construction
of superspace actions in terms of one of the analytic superfields.
There we show that scale invariance fixes the form of the
effective action uniquely, up to a coefficient. In this section,
we also study the component structure of the effective action and
confirm that it contains the $F^4/X^4$ and WZ terms.  In the
Appendices, we have collected some useful formulae for the
covariant spinor and harmonic derivatives in the analytic
coordinates, as well as for the SO(3) harmonic integrals.

\section{$\cN=4$ bi-harmonic superspace}

In this section, we define the basic structures of the $\cN=4$
bi-harmonic superspace.

\subsection{SU(2)$\times$SU(2) harmonics}

The conventional $\cN=4$\, $d=4$ superspace is described by
Minkowski space coordinates $x^m$ and Grassmann coordinates
$\theta_I^\alpha$, $\bar\theta^{I\dot\alpha}$, where $I=1,2,3,4$
is the SU(4) index. We are going to constrict a functional in the
$\cN=4$ superspace which manifestly respects only the
SU(2)$\times$SU(2)$\sim$SO(3)$\times$SO(3) subgroup of the full
SU(4)$\sim$SO(6) R-symmetry group. Therefore, it is convenient to
label Grassmann coordinates by the indices of this
SU(2)$\times$SU(2) subgroup rather than the full SU(4) group.

Let $i=1,2$ and $a=1,2$ be the indices corresponding to the two SU(2)'s.
We then represent the SU(4) index $I$ by the pair $(i,a)$:
\be
I=(i,a)=[(1,1),(1,2),(2,1),(2,2)]\,,
\label{I}
\ee
and label Grassmann variables as $\theta_{ia}^\alpha$, $\bar\theta^{ia\,\dot\alpha}$.
Here the bar indicates complex conjugation. The SU(2) indices can be raised and lowered with the SU(2) $\varepsilon$-tensors,
\be
\overline{(\theta_{ia}^\alpha)}=\bar\theta^{ia\,\dot\alpha}\,,\qquad
\theta^{ia\,\alpha}=\varepsilon^{ij}\varepsilon^{ab}\theta_{jb}^\alpha\,.
\ee
As in \cite{IN}, we now introduce two sets of harmonic variables, $u^\pm_i$ and $v^\pm_a$,
\be
u^{+i}u^-_i= v^{+a} v^-_a=1\,,\quad
u^{+i}u^+_i=u^{-i}u^-_i=0\,,\quad
v^{+a}v^+_a=v^{-a}v^-_a=0\,,
\label{u-harm}
\ee
and define the following covariant harmonic derivatives,
\bea
D^{(2,0)}=u^+_i\frac\partial{\partial u^-_i}\,,\quad
D^{(-2,0)}=u^-_i\frac\partial{\partial u^+_i}\,,\quad
S_1=[D^{(2,0)},D^{(-2,0)}]=u^+_i\frac\partial{\partial u^+_i}
-u^-_i\frac\partial{\partial u^-_i}\,,\nn\\
D^{(0,2)}=v^+_a\frac\partial{\partial v^-_a}\,,\quad
D^{(0,-2)}=v^-_a\frac\partial{\partial v^+_a}\,,\quad
S_2=[D^{(0,2)},D^{(0,-2)}]=v^+_a\frac\partial{\partial v^+_a}
-v^-_a\frac\partial{\partial v^-_a}\,.\hspace{7pt}
\label{Dh}
\eea
The operators $S_1$ and $S_2$ define U(1)
subgroups of the two SU(2)'s and measure U(1) charges of
other operators: $[S_1,D^{(s_1,s_2)}]=s_1 D^{(s_1,s_2)}$,
$[S_2,D^{(s_1,s_2)}]=s_2 D^{(s_1,s_2)}$\,.
Accordingly, we define the following bi-harmonic projections of the Grassmann variables,
\bea
\theta^{(1,1)}_\alpha=u^+_i v^+_a \theta^{ia}_\alpha\,,\quad
\theta^{(1,-1)}_\alpha=u^+_i v^-_a \theta^{ia}_\alpha\,,\quad
\theta^{(-1,1)}_\alpha=u^-_i v^+_a \theta^{ia}_\alpha\,,\quad
\theta^{(-1,-1)}_\alpha=u^-_i v^-_a \theta^{ia}_\alpha\,,
\nn\\
\bar\theta^{(1,1)}_{\dot\alpha}=u^+_i v^+_a \bar\theta^{ia}_{\dot\alpha}\,,\quad
\bar\theta^{(1,-1)}_{\dot\alpha}=u^+_i v^-_a \bar\theta^{ia}_{\dot\alpha}\,,\quad
\bar\theta^{(-1,1)}_{\dot\alpha}=u^-_i v^+_a \bar\theta^{ia}_{\dot\alpha}\,,\quad
\bar\theta^{(-1,-1)}_{\dot\alpha}=u^-_i v^-_a \bar\theta^{ia}_{\dot\alpha}\,,
\eea
where superscripts indicate the U(1) charges. To make the subsequent expressions more compact, however, we introduce a single (bold) index to represent the pairs of U(1) charges. Namely, we define
\bea
&&
\ta^{\1}_\alpha\equiv \ta^{(1,1)}_\alpha\,, \quad
\ta^{\2}_\alpha\equiv \ta^{(1,-1)}_\alpha\,, \quad
\ta^{\3}_\alpha\equiv \ta^{(-1,1)}_\alpha\,, \quad
\ta^{\4}_\alpha\equiv \ta^{(-1,-1)}_\alpha\,,\nn\\
&&
\bar\ta^{\1}_{\dot\alpha}\equiv \bar\ta^{(-1,-1)}_{\dot\alpha}\,,\quad
\bar\ta^{\2}_{\dot\alpha}\equiv \bar\ta^{(-1,1)}_{\dot\alpha}\,, \quad
\bar\ta^{\3}_{\dot\alpha}\equiv \bar\ta^{(1,-1)}_{\dot\alpha}\,, \quad
\bar\ta^{\4}_{\dot\alpha}\equiv \bar\ta^{(1,1)}_{\dot\alpha}\,.
\eea
We emphasize that these $\theta$'s have definite U(1) charges and are linear combinations of the original $\theta$'s with SU(4) indices.

Going through the same steps for the standard covariant spinor derivatives $D^I_\alpha$, $\bar D_{I\dot\alpha}$,
\be
D^I_\alpha=\frac\partial{\partial\theta_I^\alpha}+i
\bar\theta^{I\dot\alpha}\partial_{\alpha\dot\alpha}\,,\quad
\bar D_{I\dot\alpha}=-\frac\partial{\partial\bar\theta^{I\dot\alpha}}
-i\theta_I^\alpha\partial_{\alpha\dot\alpha}\,,\quad
\{D^I_\alpha,\bar D_{J\dot\alpha}
\}=-2i\delta^I_J\partial_{\alpha\dot\alpha}\,,
\ee
we define their bi-harmonic projections as
\bea
&&D_\al^\1=+\frac{\p}{\p\ta^{\1\al}}+i\tabar^{\1\ald}\p_{\al\ald}\,, \quad
\Dbar_\ald^\1=-\frac{\p}{\p\tabar^{\1\ald}}-i\ta^{\1\al}\p_{\al\ald} \nn\\
&&D_\al^\2=-\frac{\p}{\p\ta^{\2\al}}+i\tabar^{\2\ald}\p_{\al\ald}\,, \quad
\Dbar_\ald^\2=+\frac{\p}{\p\tabar^{\2\ald}}-i\ta^{\2\al}\p_{\al\ald} \nn\\
&&D_\al^\3=-\frac{\p}{\p\ta^{\3\al}}+i\tabar^{\3\ald}\p_{\al\ald}\,, \quad
\Dbar_\ald^\3=+\frac{\p}{\p\tabar^{\3\ald}}-i\ta^{\3\al}\p_{\al\ald} \nn\\
&&D_\al^\4=+\frac{\p}{\p\ta^{\4\al}}+i\tabar^{\4\ald}\p_{\al\ald}\,, \quad
\Dbar_\ald^\4=-\frac{\p}{\p\tabar^{\4\ald}}-i\ta^{\4\al}\p_{\al\ald}\,.
\label{D}
\eea
The non-trivial anticommutation relations among them are given by
\be
\{D^\1_\alpha,\bar D^\1_{\dot\alpha} \}=
\{D^\4_\alpha,\bar D^\4_{\dot\alpha}
\}=-2i\partial_{\alpha\dot\alpha}\,,\qquad
\{D^\2_\alpha,\bar D^\2_{\dot\alpha} \}=
\{D^\3_\alpha,\bar D^\3_{\dot\alpha}
\}=2i\partial_{\alpha\dot\alpha}\,.
\ee

\subsection{Tilde-conjugation}

Proper definition of conjugation in harmonic superspaces is essential for defining `real' objects. In the case at hand, complex conjugation is uniquely specified by its standard action on the SU(2) harmonics,
\be
\overline{u^{+i}}=u^-_i\,,\quad
\overline{u^+_i}=-u^{-i}\,,\quad
\overline{v^{+a}}=v^-_a\,,\quad
\overline{v^+_a}=-v^{-a}\,.
\label{cc}
\ee
However, complex conjugation turns out to be inadequate for use in analytic subspaces that we will introduce next. Instead, we will need the so-called tilde-conjugation ``$\widetilde{\phantom{m}}$'' defined as a combination of the complex conjugation ``$\overline{\phantom{m}}$'' and a special involution ``$\star$''.

In the standard $\cN=2$ harmonic superspace \cite{GIKOS,book}, the $\star$-involution is defined by
\be
(u^{+i})^\star=u^{-i}\,,\quad
(u^{+}_i)^\star=u^-_i\,,\quad
(u^{-i})^\star=-u^{-i}\,,\quad
(u^-_i)^\star=-u^+_i\,.
\label{*}
\ee
It acts only on the harmonic variables and squares to $-1$ on them. In the bi-harmonic superspace, however, we have two independent sets of harmonics variables, $u^\pm_i$ and $v^\pm_a$, and there are several ways in which the $\star$-involution can be defined. We will define it to act on $u^\pm_i$ by the rule (\ref{*}) while leaving the harmonics $v^\pm_a$ inert,
\be
(v^\pm_a)^\star= v^\pm_a\,,\qquad
(v^{\pm a})^\star= v^{\pm a}\,.
\label{**}
\ee

Combining the complex conjugation (\ref{cc}) with the $\star$-involution (\ref{*},\ref{**}), we obtain the tilde-conjugation that acts on the harmonics and Grassmann variables as follows,
\bea
&&
\widetilde{u^\pm_i }=u^{\pm i}\,,\qquad
\widetilde{u^{\pm i}}=-u^{\pm}_i\,,\nn\\
&&
\widetilde{ v^{+a} }= v^-_a\,,\quad
\widetilde{ v^{+}_a }= -v^{-a}\,,\quad
\widetilde{ v^{-a} } = - v^+_a \,,\quad
\widetilde{ v^-_a } = v^{+a}\,,
\label{tilde1}\\
&&
\widetilde{\theta^\1_\alpha}=-\bar\theta^{\3}_{\dot\alpha}\,,\quad
\widetilde{\theta^\2_\alpha}=\bar\theta^{\4}_{\dot\alpha}\,,\quad
\widetilde{\theta^\3_\alpha}=-\bar\theta^{\1}_{\dot\alpha}\,,\quad
\widetilde{\theta^\4_\alpha}=\bar\theta^{\2}_{\dot\alpha}\,,
\nn\\&&
\widetilde{\bar\theta^\1_{\dot\alpha}}=\theta^\3_\alpha\,,\quad
\widetilde{\bar\theta^\2_{\dot\alpha}}=-\theta^\4_\alpha\,,\quad
\widetilde{\bar\theta^\3_{\dot\alpha}}=\theta^\1_\alpha\,,\quad
\widetilde{\bar\theta^\4_{\dot\alpha}}=-\theta^\2_\alpha\,.
\label{tilde2}
\eea
As we will see later, this is exactly the conjugation that will allow us to introduce real actions in the $\cN=4$ bi-harmonic superspace.

\subsection{Analytic subspaces}

By definition, an analytic subspace in the full $\cN=4$ superspace must (i) depend on half of Grassmann variables of the full superspace, (ii) be closed under $\cN=4$ supersymmetry, (iii) contain an equal number of $\theta$ and $\bar\theta$ variables. (For comparison, chiral subspaces depend on $\theta$ variables only.) To construct such subspaces, we pass from standard bosonic coordinates $x^m$ to the analytic ones,
\be
x_{A}^m=x^m
+a_1(i\ta^\1\si^m\tabar^\1)+a_2(i\ta^\2\si^m\tabar^\2)
+a_3(i\ta^\3\si^m\tabar^\3)
+a_4(i\ta^\4\si^m\tabar^\4)
\,,
\ee
where $a_k=\pm1$. Choosing $+1$ or $-1$ for each $a_k$ fixes the
corresponding analytic subspace. In fact, there are six such analytic
subspaces, corresponding to six different ways of choosing two out of
four Grassmann variables:
\bea
\ba[b]{c|c|c|c}
 & \text{coordinates} & (a_1,a_2,a_3,a_4) & \text{short derivatives} \\
 \hline\rule{0pt}{15pt}
A_1 & \{x^m_{A_1}, \theta^\1_\alpha, \bar\theta^\2_{\dot\alpha} ,\bar\theta^\3_{\dot\alpha}, \theta^\4_\alpha,u,v \} &
 (+,+,+,+) & \bar D^\1_{\dot\alpha}, D^\2_\alpha, D^\3_\alpha, \bar D^\4_{\dot\alpha}\\[3pt]
\bar{A}_1 & \{x^m_{\bar A_1}, \bar\theta^\1_{\dot\alpha}, \theta^\2_\alpha, \theta^\3_\alpha,\bar\theta^\4_{\dot\alpha},u,v \} &
 (-,-,-,-) & D^\1_\alpha, \bar D^\2_{\dot\alpha}, \bar D^\3_{\dot\alpha},D^\4_\alpha\\[3pt]
A_2 & \{x^m_{A_2}, \theta^\1_\alpha, \theta^\2_\alpha,
\bar\theta^\3_{\dot\alpha}, \bar\theta^\4_{\dot\alpha} ,u,v \} &
 (+,-,+,-) & \bar D^\1_{\dot\alpha}, \bar D^\2_{\dot\alpha}, D^\3_\alpha,D^\4_\alpha\\[3pt]
\bar{A}_2 & \{x^m_{\bar A_2},
 \bar\theta^\1_{\dot\alpha} ,\bar\theta^\2_{\dot\alpha},
\theta^\3_\alpha, \theta^\4_\alpha,
 u,v \} &
 (-,+,-,+) & D^\1_\alpha, D^\2_\alpha, \bar D^\3_{\dot\alpha},\bar D^\4_{\dot\alpha}\\[3pt]
A_3 & \{x^m_{A_3}, \theta^\1_\alpha,\bar\theta^\2_{\dot\alpha}, \theta^\3_\alpha,
 \bar\theta^\4_{\dot\alpha} ,u,v \} &
 (+,+,-,-) & \bar D^\1_{\dot\alpha}, D^\2_\alpha, \bar D^\3_{\dot\alpha},D^\4_\alpha\\[3pt]
\bar{A}_3 & \{x^m_{\bar A_3}, \bar\theta^\1_{\dot\alpha} , \theta^\2_\alpha,
\bar\theta^\3_{\dot\alpha}, \theta^\4_\alpha,u,v \} &
 (-,-,+,+) & D^\1_\alpha, \bar D^\2_{\dot\alpha}, D^\3_\alpha,\bar D^\4_{\dot\alpha}
\label{table}
\ea
\eea
In the last column, we listed covariant spinor derivatives that
become `short' in the corresponding coordinates. They differentiate
along those Grassmann directions which are \emph{orthogonal} to the
corresponding analytic subspace.

Under the tilde-conjugation (\ref{tilde1},\ref{tilde2}), the analytic
subspaces $A_1$, $\bar A_1$, $A_2$ and $\bar A_2$ are \emph{real},
whereas $A_3$ and $\bar A_3$ transform into each other.~\footnote{The
subspaces $A_3$ and $\bar A_3$ would be real under \emph{modified}
tilde-conjugation where the behavior of the harmonics $u$ and $v$
 with respect to the $\star$-involution (\ref{*},\ref{**}) is
 reversed.} In what follows, we will explicitly consider only the
 subspaces $A_1$ and $A_2$. The expressions for the covariant spinor
 (\ref{D}) and harmonic (\ref{Dh}) derivatives in these subspaces
 are given in Appendix A.

The remaining analytic subspaces could be treated similarly, but
this would not produce any qualitatively new results.

\section{$\cN=4$ SYM in bi-harmonic superspace}

In this section, we start with the standard $\cN=4$ supergauge field strength and use it to define six different harmonic superfields, each of which \emph{independently} can be used to describe the on-shell $\cN=4$ SYM multiplet.

\subsection{Constraints on the $\cN=4$ gauge superfield strength}

In the conventional $\cN=4$ superspace, the $\cN=4$ superfield strength is described by an antisymmetric SU(4) tensor, $W^{IJ}=-W^{JI}$, subject to the following constraints \cite{S1,S2},
\bea
&&\overline{W^{IJ}} \equiv \bar W_{IJ}=\frac12\varepsilon_{IJKL}W^{KL}\,,
\label{c3}\\
&& D^I_\alpha W^{JK}+D^J_\alpha W^{IK}=0\,,\label{c2}\\
&&\bar D_{I\dot\alpha} W^{JK}=\frac13(\delta_I^J\bar D_{L\dot\alpha} W^{LK}
 -\delta_I^K\bar D_{L\dot\alpha}W^{LJ})\,.
\label{c1}
\eea
Writing the SU(4) indices as pairs of SU(2) ones, as in (\ref{I}), we find
\be
W^{IJ} \equiv W^{ia,jb}=\varepsilon^{ij}W^{ab}+\varepsilon^{ab}W^{ij}\,,
\ee
so that the superfield strength $W^{IJ}$ becomes represented by a pair of \emph{symmetric} SU(2) tensors: $W^{ab}=W^{ba}$ and $W^{ij}=W^{ji}$. The constraints (\ref{c3})--(\ref{c1}) can be readily rewritten for these tensors. In particular, using the identity
\be
\varepsilon_{IJKL} \equiv
\varepsilon_{ia,jb,kc,ld}=
\varepsilon_{il}\varepsilon_{jk}\varepsilon_{ab}\varepsilon_{cd}
-\varepsilon_{ij}\varepsilon_{kl}\varepsilon_{ad}\varepsilon_{bc}\,,
\ee
we find that (\ref{c3}) corresponds to the following reality properties of the SU(2) tensors,
\be
\overline{W^{ij}} \equiv \bar W_{ij}=W_{ij}\,,\qquad
\overline{W^{ab}} \equiv \bar W_{ab}=-W_{ab}\,.
\label{Wconj}
\ee
The constraint (\ref{c2}) is equivalent to the following set of constraints,
\be
D^{a(i}_\alpha W^{jk)}=0\,,\quad
D^{i(a}_\alpha W^{bc)}=0\,,\quad
D^{ka}_\alpha W^i_k+D^{ic}_\alpha W^a_c=0\,.
\label{cc2}
\ee
The constraint (\ref{c1}) leads to equations conjugate to these,~\footnote{
It is straightforward to show that (\ref{c3}) and (\ref{c2}) together imply (\ref{c1}).
}
\be
\bar D^{a(i}_{\dot\alpha} W^{jk)}=0\,,\quad
\bar D^{i(a}_{\dot\alpha }W^{bc)}=0\,,\quad
\bar D^{ka}_{\dot\alpha} W^i_k-\bar D^{ic}_{\dot\alpha} W^a_c=0\,.
\label{cc1}
\ee
We conclude that equations (\ref{Wconj}), (\ref{cc2}) and (\ref{cc1}) are equivalent to the $\cN=4$ supergauge constraints (\ref{c3}), (\ref{c2}) and (\ref{c1}).

\subsection{Harmonic projections and solutions to the constraints}

Now we introduce the harmonic projections for the superfields $W^{ij}$ and $W^{ab}$,
\bea
&&
W=u^+_i u^-_j W^{ij}-v^+_a v^-_b W^{ab}\,,\qquad
W'=u^+_i u^-_j W^{ij}+v^+_a v^-_b W^{ab}\,,\label{35}\\
&&W^{(2,0)}=u^+_i u^+_j W^{ij}\,,\qquad
W^{(-2,0)}=u^-_i u^-_j W^{ij}\,,\label{36}\\&&
W^{(0,2)}=v^+_a v^+_b W^{ab}\,,\qquad
W^{(0,-2)}=v^-_a v^-_b W^{ab}\,.
\label{37}
\eea
According to the conjugation rules (\ref{tilde1}) and (\ref{Wconj}), these harmonic projections have the following reality properties,
\be
\widetilde{W}={W}\,,\quad
\widetilde{{W}'}={W}'\,,\quad
\widetilde{W^{(\pm 2,0)}}=W^{(\pm 2,0)}\,,\quad
\widetilde{W^{(0,\pm2)}}=- W^{(0,\mp2)}\,.
\label{conj}
\ee
Contracting the constraints (\ref{cc2}) and (\ref{cc1}) with various combinations of harmonic variables, we find the following first-order differential constraints,
\bea
&&
\{\bar D^\1_{\dot\alpha}, D^\2_\alpha ,D^\3_\alpha,
\bar D^\4_{\dot\alpha}\}{ W}=0\,,
\label{Wan}\\
&&
\{ D^\1_\alpha,
\bar D^\2_{\dot\alpha},\bar D^\3_{\dot\alpha},D^\4_\alpha\} {W}'=0\,,\\
&&
\{\bar D^\1_{\dot\alpha} ,
\bar D^\2_{\dot\alpha},D^\3_\alpha ,D^\4_\alpha\} W^{(2,0)}=0\,,
\label{W20an}\\
&&
\{ D^\1_\alpha,
D^\2_\alpha,
\bar D^\3_{\dot\alpha} ,
\bar D^\4_{\dot\alpha}\}W^{(-2,0)}=0\,,\\
&&
\{\bar D^\1_{\dot\alpha} ,
 D^\2_\alpha ,\bar D^\3_{\dot\alpha},D^\4_\alpha \} W^{(0,2)}=0\,,\\
&&
\{ D^\1_\alpha  ,
\bar D^\2_{\dot\alpha},D^\3_\alpha ,
\bar D^\4_{\dot\alpha}\} W^{(0,-2)}=0\,,
\eea
as well as additional first-order differential constraints that \emph{mix} different harmonic projections. However, the mixing can be removed at the price of generating the following \emph{second}-order differential constraints,
\bea
&&
\{(D^\1)^2,(\bar D^\2)^2,(\bar D^\3)^2,(D^\4)^2,(D^\1 D^\4),
(\bar D^\2\bar D^\3)\}{ W}=0\,,
\label{Wlin}\\
&&
\{(\bar D^\1)^2,(D^\2)^2, (D^\3)^2,(\bar D^\4)^2,(\bar D^\1\bar D^\4), (D^\2 D^\3)
\}{ W}'=0\,,\\
&&
\{
(D^\1)^2,(D^\2)^2,(\bar D^\3)^2,(\bar D^\4)^2,(D^\1 D^\2),
(\bar D^\3\bar D^\4)\}  W^{(2,0)}=0\,,\label{W20lin}\\
&&
\{(\bar D^\1)^2,(\bar D^\2)^2,(D^\3)^2,(D^\4)^2,
(\bar D^\1 \bar D^\2),(D^\3 D^\4)\} W^{(-2,0)}=0\,,\\
&&
\{(D^\1)^2, (\bar D^\2)^2,(D^\3)^2,(\bar D^\4)^2,
(D^\1 D^\3),
(\bar D^\2\bar D^\4)\}
 W^{(0,2)}=0\,,\\
&&
\{(\bar D^\1)^2,(D^\2)^2,(\bar D^\3)^2,(D^\4)^2,
(\bar D^\1\bar D^\3),(D^\2 D^\4)\}
 W^{(0,-2)}=0\,.
\eea
Finally, there are many differential relations for the superfield strengths involving covariant \emph{harmonic} derivatives, which follow from the definitions (\ref{35})--(\ref{37}). The basic constraints in this set are
\bea
&&D^{(2,0)}D^{(2,0)}{ W}=D^{(0,2)}D^{(0,2)}{ W}=
D^{(2,0)}D^{(0,2)}{ W}=0\,,
\label{Wharm}\\
&&D^{(2,0)}D^{(2,0)}{ W}'=D^{(0,2)}D^{(0,2)}{ W}'=
D^{(2,0)}D^{(0,2)}{ W}'=0\,,\\
&&D^{(2,0)}W^{(2,0)}=D^{(0,2)}W^{(2,0)}=D^{(0,-2)}W^{(2,0)}=0\,,
\label{W20harm}
\\
&&D^{(-2,0)}W^{(-2,0)}=D^{(0,2)}W^{(-2,0)}=D^{(0,-2)}W^{(-2,0)}=0\,,
\\
&&D^{(0,2)}W^{(0,2)}=D^{(2,0)}W^{(0,2)}=D^{(-2,0)}W^{(0,2)}=0\,,
\\
&&D^{(0,-2)}W^{(0,-2)}=D^{(2,0)}W^{(0,-2)}=D^{(-2,0)}W^{(0,-2)}=0\,.
\label{last}
\eea
Note that the constraints for the chargeless superfields $W$ and $W^\prime$ are \emph{quadratic} in the harmonic derivatives, whereas those for the charged superfields $W^{(\pm2,0)}$ and $W^{(0,\pm2)}$ are \emph{linear} in these derivatives.

Our claim now is that equations (\ref{conj})--(\ref{last}) form the complete set of constraints that eliminate all the auxiliary fields in the component expansions of $ W$, ${ W}'$, $W^{(\pm2,0)}$, $W^{(0,\pm2)}$, reducing each of them to the \emph{on-shell} $\cN=4$ supergauge multiplet.  We will demonstrate this next by giving explicit solutions of these constraints for the two inequivalent cases: the chargeless superfield $W$ and the charged superfield $W^{(2,0)}$. The other four cases yield qualitatively similar results.

\subsubsection{Chargeless superfield}
\label{chargeless}

The chargeless superfield $W$ obeys the constraints (\ref{conj}), (\ref{Wan}), (\ref{Wlin}), and (\ref{Wharm}).  The constraints (\ref{Wan}) are nothing but the analyticity conditions. They are solved by passing to the coordinates of the analytic subspace $A_1$ given in (\ref{table}),
\be
W=W(x^m_{A_1},\theta^\1_\alpha,\bar\theta^\2_{\dot\alpha},\bar\theta^\3_{\dot\alpha},
\theta^\4_\alpha,u,v)\,.
\label{3.32}
\ee
Although this superfield depends on half of the Grassmann variables of the $\cN=4$ superspace, its component field decomposition is still quite long. However, it becomes shorter upon taking into account the linearity conditions (\ref{Wlin}).  To the remaining components we have to apply the constraints with the covariant harmonic derivatives (\ref{Wharm}). Note that after passing to the analytic coordinates these constraints become dynamical because the covariant harmonic derivatives (\ref{HA1}) involve space-time derivatives. Taking into account all these equations, we obtain the following component field decomposition for $W$,
\bea
W&=&\omega+u^+_i u^-_j \phi^{ij}+v^+_a v^-_b i \varphi^{ab}\nn\\&&
+\theta^{\1\alpha}\psi_\alpha^{ia}u^-_i v^-_a
-\theta^{\4\alpha}\psi_\alpha^{ia}u^+_i v^+_a
+\bar\theta^{\2}_{\dot\alpha}\bar\psi^{\dot\alpha\, ia}u^+_i
v^-_a
-\bar\theta^{\3}_{\dot\alpha}\bar\psi^{\dot\alpha\, ia}
u^-_i v^+_a
\nn\\&&
+\frac1{\sqrt2}(\theta^\1_\alpha \theta^\4_\beta
 \sigma^{m\alpha}{}_{\dot\alpha}\sigma^{n\beta\dot\alpha}
+\bar\theta^\3_{\dot\alpha}\bar\theta^\2_{\dot\beta}\sigma^{m\dot\alpha}{}_\alpha
\sigma^{n\alpha\dot\beta})F_{mn}\nn\\&&
+2\theta^{\1\alpha}\bar\theta^{\2\dot\alpha}
 \partial_{\alpha\dot\alpha}\varphi^{ab} v^-_a v^-_b
+2\theta^{\4\alpha}\bar\theta^{\3\dot\alpha}
 \partial_{\alpha\dot\alpha}\varphi^{ab} v^+_a v^+_b
\nn\\&&
-2i \theta^{\4\alpha}\bar\theta^{\2\dot\alpha}
 \partial_{\alpha\dot\alpha} \phi^{ij}u^+_i u^+_j
-2i \theta^{\1\alpha}\bar\theta^{\3\dot\alpha}
 \partial_{\alpha\dot\alpha}\phi^{ij} u^-_i u^-_j
 \nn\\&&
+2i\theta^{\4\alpha}\theta^{\1\beta}\bar\theta^{\3\dot\beta}
 \partial_{\beta\dot\beta}\psi^{ia}_\alpha u^-_i v^+_a
-2i\theta^{\1\alpha}\theta^{\4\beta}\bar\theta^{\2\dot\beta}
\partial_{\beta\dot\beta}\psi^{ia}_\alpha u^+_i v^-_a
\nn\\&&
+2i\bar\theta^{\3\dot\alpha}\theta^{\4\beta}\bar\theta^{\2\dot\beta}
\partial_{\beta\dot\beta}\bar\psi^{ia}_{\dot\alpha} u^+_i v^+_a
-2i\bar\theta^{\2\dot\alpha}\theta^{\1\beta}\bar\theta^{\3\dot\beta}
\partial_{\beta\dot\beta}\bar\psi^{ia}_{\dot\alpha} u^-_i v^-_a
\nn\\&&
+4\theta^{\1\alpha}\theta^{\4\beta}\bar\theta^{\3\dot\alpha}
\bar\theta^{\2\dot\beta}\partial_{\alpha\dot\alpha}\partial_{\beta\dot\beta}
(u^+_i u^-_j \phi^{ij}-v^+_a v^-_b i \varphi^{ab})\,.
\label{W}
\eea
Here $\omega$ is a \emph{constant}, $\phi^{ij}=\phi^{(ij)}$ and $\varphi^{ab}=\varphi^{(ab)}$ are two triplets of scalar fields, $\psi^{ia}_\alpha$ are four Weyl spinors and $F_{mn}$ is the Maxwell field strength.  These fields obey their classical equations of motion,
\be
\square \phi^{ij}=\square \varphi^{ab}=0\,,\quad
\partial^{\alpha\dot\alpha}\psi^{ia}_\alpha=0\,,\quad
\partial^m F_{mn}=0\,.
\label{eom}
\ee
No auxiliary field components remain in $W$ as they all have vanished under the constraints (\ref{Wan}),
(\ref{Wlin}) and (\ref{Wharm}).

Let us comment on the constant $\omega$ in (\ref{W}). This constant would never have arisen if we started with the component form of $W^{IJ}$ that solves (\ref{c3})--(\ref{c1}), constructed $W$ using (\ref{35}) and made the transformation to analytic coordinates $A_1$. However, we instead considered $W$ to be \emph{defined} by the constraints (\ref{conj}), (\ref{Wan}), (\ref{Wlin}), and (\ref{Wharm}). These constraints were, indeed, sufficient to properly restrict the component degrees of freedom, except for the residual appearance of this extra constant parameter.

We will set $\omega$ to zero by insisting that $W$ transforms linearly under scale transformations, with a constant parameter $\lambda$,
\be
\delta W=\lambda W \quad \Rightarrow \quad \omega=0\,.
\ee
This requirement is particularly natural for the purposes of the next section, where we will construct the superconformal effective action in the $\cN=4$ supergauge theory.

\subsubsection{Charged superfield}
\label{charged}

Now let us briefly consider the charged superfield $W^{(2,0)}$ subject to the differential constraints (\ref{W20an}), (\ref{W20lin}), (\ref{W20harm}) and the reality condition (\ref{conj}).  The analyticity constraints (\ref{W20an}) are solved by passing to the analytic coordinates $A_2$ given in (\ref{table}),
\be
W^{(2,0)}=W^{(2,0)}(x^m_{A_2}, \theta^\1_\alpha, \theta^\2_\alpha,
\bar\theta^\3_{\dot\alpha}, \bar\theta^\4_{\dot\alpha} ,u,v)\,.
\label{3.36}
\ee
Using the linearity conditions (\ref{W20lin}) and the `harmonic shortness' constraints (\ref{W20harm}), we obtain
\bea
W^{(2,0)}&=&\phi^{ij}u^+_i u^+_j\nn\\&&
+\theta^{\1\alpha}\psi^{ia}_\alpha u^+_i v^-_a
-\theta^{\2\alpha}\psi^{ia}_\alpha u^+_i v^+_a
+\bar\theta^\4_{\dot\alpha}\bar\psi^{ia\,\dot\alpha} u^+_i v^-_a
-\bar\theta^\3_{\dot\alpha}\bar\psi^{ia\,\dot\alpha} u^+_i v^+_a
\nn\\&&
+\frac 1{\sqrt2}(\theta^\1_\alpha\theta^\2_\beta
 \sigma^{m\alpha}{}_{\dot\alpha}\sigma^{n\beta\dot\alpha}
 +\bar\theta^\3_{\dot\alpha}\bar\theta^\4_{\dot\beta}
 \sigma^{m\dot\alpha}{}_\alpha \sigma^{n\alpha\dot\beta})F_{mn}
\nn\\&&
+2\theta^{\2\alpha}\bar\theta^{\3\dot\alpha}\partial_{\alpha\dot\alpha}
 \varphi^{ab}v^+_a v^+_b
+2i\theta^{\2\alpha}\bar\theta^{\4\dot\alpha}
 \partial_{\alpha\dot\alpha}(\phi^{ij}u^+_i u^-_j+i\varphi^{ab}v^+_a v^-_b)
\nn\\&&
+2\theta^{\1\alpha}\bar\theta^{\4\dot\alpha}\partial_{\alpha\dot\alpha}
 \varphi^{ab}v^-_a v^-_b
+2i\theta^{\1\alpha}\bar\theta^{\3\dot\alpha}\partial_{\alpha\dot\alpha}
 (-\phi^{ij}u^+_i u^-_j +i\varphi^{ab}v^+_a v^-_b)
\nn\\&&
+2i\theta^{\1\alpha}\theta^{\2\beta}\bar\theta^{\4\dot\beta}
 \partial_{\beta\dot\beta}\psi^{ia}_\alpha u^-_i v^-_a
+2i\theta^{\2\alpha}\theta^{\1\beta}\bar\theta^{\3\dot\beta}
 \partial_{\beta\dot\beta}\psi^{ia}_\alpha u^-_i v^+_a
\nn\\&&
+2i\bar\theta^{\3\dot\alpha}\theta^{\2\beta}\bar\theta^{\4\dot\beta}
 \partial_{\beta\dot\beta}\bar\psi^{ia}_{\dot\alpha} u^-_i v^+_a
+2i\bar\theta^{\4\dot\alpha}\theta^{\1\beta}\bar\theta^{\3\dot\beta}
 \partial_{\beta\dot\beta}\bar\psi^{ia}_{\dot\alpha} u^-_i v^-_a
\nn\\&&
-4\theta^{\1\alpha}\theta^{\2\beta}\bar\theta^{\3\dot\alpha}
 \bar\theta^{\4\dot\beta}
 \partial_{\alpha\dot\alpha}\partial_{\beta\dot\beta}
 \phi^{ij}u^-_i u^-_j\,.
\label{W20}
\eea
All auxiliary field components have vanished, whereas the physical
components are required to satisfy the free equations of motion
(\ref{eom}). In this case, unlike (\ref{W}), no extra constant
parameter appears in the solution. Note, however, that $W^{(2,0)}$
involves only half of the scalars, $\phi^{ij}$, undifferentiated;
the remaining scalars, $\varphi^{ab}$, appear only with the derivatives
acting on them. This limits the range of applications of the charged
superfield, and makes $W$ the preferred choice for the description
of the $\cN=4$ SYM multiplet.

\section{$\cN=4$ SYM effective action}
\label{Sec4}

In this section, we find the bi-harmonic superspace form of the $\cN=4$ SYM effective action on the Coulomb branch, and confirm that it correctly reproduces the $F^4/X^4$ and WZ terms.

\subsection{The superfield effective action}

The simplest $\cN=4$ superspace action for the superfield $W$ is given by
\be
\Gamma=\int d\zeta du dv\,H(W)\,,
\label{G}
\ee
where $H(W)$ is some function of $W$ without derivatives.  The integration goes over the analytic superspace $A_1$ given in (\ref{table}) with the analytic measure defined so that
\be
d\zeta = d^4x d^8\theta\,,\qquad
\int d^8\theta
\,(\theta^\1)^2(\theta^\4)^2(\bar\theta^\2)^2(\bar\theta^\3)^2=1\,.
\label{measure}
\ee
We use the standard definition for the harmonic integrals \cite{GIKOS,book},
\be
\int du\,1=1\,,\qquad
\int du(\mbox{non-singlet SU(2) irreducible representation})=0\,,
\label{u-int}
\ee
and similarly for the integration over $dv$. We point out that the
function $H(W)$ must have zero U(1) charges since the integration
measure $d\zeta$ in the analytic superspace $A_1$ is chargeless.

Note that the integration measure (\ref{measure}) yields eight Grassmann
derivatives, or, equivalently, four space-time ones.  Therefore, we expect
that the action (\ref{G}) with a particular $H$ describes the
four-derivative term in the $\cN=4$ low-energy effective action, and that
this term is the leading one in the derivative expansion. We will now
determine the appropriate function $H$ by requiring scale invariance of
the action (\ref{G}), in exactly the same way as we did in \cite{BS}.

As the measure $d\zeta$ is dimensionless, the function $H(W)$ must also be dimensionless. Recalling that $W$ has mass dimension one, we are forced to introduce a parameter $\Lambda$ such that $W/\Lambda$ is dimensionless, and take $H=H(W/\Lambda)$. However, the dependence on $\Lambda$ should disappear after the integration over Grassmann variables. This uniquely fixes this function in the form
\be
H=c\ln\frac W\Lambda\,,
\label{H}
\ee
with some coefficient $c$. Indeed, rescaling $W$ then shifts the integrand in (\ref{G}) by a constant which gives zero under the integral over the Grassmann variables.

In the following, we will demonstrate that the action (\ref{G})
with the function $H$ given by (\ref{H}) does contain the known
bosonic terms in the $\cN=4$ SYM effective action.
The parameter $\Lambda$ drops out in the final results, and to simplify the following expressions we formally set $\Lambda=1$ from now on.

\subsection{Bosonic terms and SO(3) harmonics}

In the bosonic part of $W$ in (\ref{W}), that is after
setting $\psi_\al^{i a}=\bar\psi_{\dot\al}^{i a}=0$, only the
following harmonic monomials appear: $u^+_i u^+_j$, $u^-_i u^-_j$, $u^+_{(i}u^-_{j)}$,
and $v^+_a v^+_b$, $v^-_a v^-_b$, $v^+_{(a}v^-_{b)}$.
For computational reasons, it is convenient to rewrite
these SU(2) monomials in terms of SO(3) harmonics
$U^1_{A'}$, $U^2_{A'}$, $U^3_{A'}$ and $V^1_A$, $V^2_A$, $V^3_A$,
\bea
&&
V^1_A= i \gamma_A^{ab} u^+_a u^-_b\,,\quad
V^2_A=\frac12\gamma_A^{ab}(u^+_a u^+_b + u^-_a u^-_b)
\,,\quad
V^3_A=\frac i2\gamma_A^{ab}(u^+_a u^+_b - u^-_a u^-_b)\,,
\nn\\&&
U^1_{A'}= i \gamma_{A'}^{ij} u^+_i u^-_j\,,\quad
U^2_{A'}=\frac12\gamma_{A'}^{ij}(u^+_i u^+_j + u^-_i u^-_j)
\,,\quad
U^3_{A'}=\frac i2\gamma_{A'}^{ij}(u^+_i u^+_j - u^-_i u^-_j)\,,
\hspace{30pt}
\label{B2}
\eea
where $\gamma^A_{ab}$, $\gamma^{A'}_{ij}$ are two copies of SO(3)
gamma-matrices,
\be
\gamma^A_{ab}\gamma^{B\,bc}+\gamma^B_{ab}\gamma^{A\,bc}=
2\delta^{AB}\delta_a^c\,,\qquad
\gamma^{A'}_{ij}\gamma^{{B'}\,jk}+\gamma^{B'}_{ij}\gamma^{{A'}\,jk}=
2\delta^{{A'}{B'}}\delta_i^k\,.
\label{gamma}
\ee
Using (\ref{u-harm}), (\ref{cc}) and (\ref{gamma}) it is straightforward to
check that the objects (\ref{B2}) are real under usual complex
conjugation and obey standard properties of SO(3) matrices,
\be
\begin{array}{lll}
U^{A'}_{B'} U^{C'}_{B'}=\delta^{{A'}{C'}}\,,\quad&
\varepsilon^{A'B'C'}U^1_{A'}U^2_{B'}U^3_{C'}=1\,,\quad&
\overline{U^{B'}_{A'}}=U^{B'}_{A'}\,,\\
V^A_B V^C_B=\delta^{AC}\,,&
\varepsilon^{ABC}V^1_{A}V^2_{B}V^3_{C}=1\,,&
\overline{V^A_B}=V^A_B\,.
\end{array}
\label{SO3harm}
\ee

In terms of the SO(3)-harmonics (\ref{B2}) the bosonic components
of the superfield (\ref{W}) can be written as
\bea
W &=& \varphi^A V^1_A-i \phi^{A'} U^1_{A'}
+\frac1{\sqrt2}(\theta^\1_\alpha \theta^\4_\beta
 \sigma^{m\alpha}{}_{\dot\alpha}\sigma^{n\beta\dot\alpha}
+\bar\theta^\3_{\dot\alpha}\bar\theta^\2_{\dot\beta}\sigma^{m\dot\alpha}{}_\alpha
\sigma^{n\alpha\dot\beta})F_{mn} \nn\\&&
+2\theta^{\1\alpha}\bar\theta^{\2\dot\alpha}
 \partial_{\alpha\dot\alpha}\varphi^A (V^2_A+iV^3_A)
+2\theta^{\4\alpha}\bar\theta^{\3\dot\alpha}
 \partial_{\alpha\dot\alpha}\varphi^A (V^2_A-i V^3_A)
\nn\\&&
-2i \theta^{\4\alpha}\bar\theta^{\2\dot\alpha}
 \partial_{\alpha\dot\alpha} \phi^{A'}(U^2_{A'}-iU^3_{A'})
-2i \theta^{\1\alpha}\bar\theta^{\3\dot\alpha}
 \partial_{\alpha\dot\alpha}\phi^{A'}(U^2_{A'}+i U^3_{A'})
 \nn\\&&
-4\theta^{\1\alpha}\theta^{\4\beta}\bar\theta^{\3\dot\alpha}
\bar\theta^{\2\dot\beta}\partial_{\alpha\dot\alpha}\partial_{\beta\dot\beta}
(V^1_A\varphi^A+i U^1_{A'}\phi^{A'})
\,,
\label{Wbos}
\eea
where we have defined the SO(3) triplets of the scalars as
\be
\varphi^A=\frac12\gamma^A_{ab}\varphi^{ab}\,,\qquad
\phi^{A'}=\frac12\gamma^{A'}_{ij}\phi^{ij}\,.
\ee
In what follows,
we will use the expression (\ref{Wbos}) to analyze the
component structure of the action (\ref{G}) in the bosonic
sector.

\subsection{The $F^4/X^4$ term}

To identify the $F^4/X^4$ term in the effective action, we neglect terms with derivatives of scalar fields in (\ref{Wbos}), so that
\be
W= \varphi^A V^1_A-i \phi^{A'} U^1_{A'}
+\frac1{\sqrt2}(\theta^\1_\alpha \theta^\4_\beta
 \sigma^{m\alpha}{}_{\dot\alpha}\sigma^{n\beta\dot\alpha}
+\bar\theta^\3_{\dot\alpha}\bar\theta^\2_{\dot\beta}\sigma^{m\dot\alpha}{}_\alpha
\sigma^{n\alpha\dot\beta})F_{mn}
\,.
\label{WW}
\ee
Substituting (\ref{WW}) into (\ref{G}) and integrating over the Grassmann variables by the rule (\ref{measure}), we find
\be
\Gamma_{F^4}=\frac14\int d^4x dU dV\,H^{(4)}(\varphi^A V^1_A-i\phi^{A'} U^1_{A'})
\left[F_{mn}F^{nk}F_{kl}F^{lm}-\frac14(F_{pq}F^{pq})^2\right]\,.
\label{GF}
\ee
Here we have applied the standard identity for the trace of four sigma-matrices,
\be
\tr\,\tilde\sigma^m\sigma^n\tilde\sigma^p
\sigma^q=-2i\varepsilon^{mnpq}
+2(\eta^{mn}\eta^{pq}+\eta^{np}\eta^{mq}-\eta^{mp}\eta^{nq})
\,.
\label{tr-sigma}
\ee
Choosing now the function $H$ as in (\ref{H}), we expand it in the Taylor series over $i\phi^{A'}U^1_{A'}$,
\bea
H^{(4)}(\varphi^A V^1_A-i\phi^{A'} U^1_{A'})
&=&\sum_{n=0}^\infty \frac1{n!}H^{(n+4)}(\varphi^A V^1_A) (-i\phi^{A'}U^1_{A'})^n
\nn\\&=&
c\sum_{n=0}^\infty \frac{(-1)^{n+1}(n+3)!}{n!}
\frac{(-i\phi^{A'} U^1_{A'})^n}{(\varphi^A V^1_A)^{n+4}}\,.
\label{ser}
\eea
Here $H^{(n)}$ stands for the $n$'th derivative of the function
$H$ with respect to its argument.
Substituting this decomposition into (\ref{GF}) and computing the harmonic integral over $dU$ using (\ref{hint1}), we obtain
\bea
\Gamma_{F^4}&=&-\frac c4\int d^4x dV\,[F_{mn}F^{nk}F_{kl}F^{lm}-\frac14(F_{pq}F^{pq})^2]
\sum_{n=0}^\infty (2n+2)(2n+3)
\frac{(-\phi^{A'}\phi^{A'})^{n}}{(\varphi^A V^1_A)^{2n+4}}
\nn\\&=&
\frac c2\int d^4x dV\,[F_{mn}F^{nk}F_{kl}F^{lm}-\frac14(F_{pq}F^{pq})^2]
\frac{\phi^{A'}\phi^{A'}-3(\varphi^A V^1_A)^2}{[\phi^{D'}\phi^{D'}+(\varphi^D V^1_D)^2]^3}
\,.
\label{4.9}
\eea
It is interesting to note that the series
in the first line in (\ref{4.9}) reduced to the concise analytical expression
given in the second line. This allows us to expand the
expression in the second line in (\ref{4.9}) in a series over
another argument, $\varphi^A V^1_A$,
and compute the harmonic integral over $dV$,
\bea
\Gamma_{F^4}&=&\frac c2\int d^4x dV
\frac{F_{mn}F^{nk}F_{kl}F^{lm}-\frac14(F_{pq}F^{pq})^2}{(\phi^{D'}\phi^{D'})^2}
\sum_{n=0}^\infty (-1)^n(2n+1)(n+1)
\frac{(\varphi^A V^1_A)^{2n}}{(\phi^{A'}\phi^{A'})^n}
\nn\\&=&
\frac c2\int d^4x \frac{F_{mn}F^{nk}F_{kl}F^{lm}-\frac14(F_{pq}F^{pq})^2}{(\phi^{D'}\phi^{D'})^2}
\sum_{n=0}^\infty (-1)^n(n+1)\left(
\frac{\varphi^A \varphi^A}{\phi^{A'}\phi^{A'}}
\right)^n .
\eea
This series can be easily summed up, and we find the following result
\be
\label{F4X4}
\Gamma_{F^4}=\frac c2\int d^4x
\frac{F_{mn}F^{nk}F_{kl}F^{lm}-\frac14(F_{pq}F^{pq})^2}{(\phi^{A'}\phi^{A'}
+\varphi^A\varphi^A)^2}\,.
\ee
Note that the scalar fields in the denominator appear in an SO(6)-invariant form.

\subsection{The WZ term}

To identify the Wess-Zumino term in the effective action, we omit the Maxwell field strength from the expansion in (\ref{Wbos}), so that now
\bea
W&=&V^1_A\varphi^A-i U^1_{A'} \phi^{A'}
\nn\\&&
+2\theta^{\1\alpha}\bar\theta^{\2\dot\alpha}
 \partial_{\alpha\dot\alpha}\varphi^A (V^2_A+iV^3_A)
+2\theta^{\4\alpha}\bar\theta^{\3\dot\alpha}
 \partial_{\alpha\dot\alpha}\varphi^A (V^2_A-i V^3_A)
\nn\\&&
-2i \theta^{\4\alpha}\bar\theta^{\2\dot\alpha}
 \partial_{\alpha\dot\alpha} \phi^{A'}(U^2_{A'}-iU^3_{A'})
-2i \theta^{\1\alpha}\bar\theta^{\3\dot\alpha}
 \partial_{\alpha\dot\alpha}\phi^{A'}(U^2_{A'}+i U^3_{A'})
 \nn\\&&
-4\theta^{\1\alpha}\theta^{\4\beta}\bar\theta^{\3\dot\alpha}
\bar\theta^{\2\dot\beta}\partial_{\alpha\dot\alpha}\partial_{\beta\dot\beta}
(V^1_A\varphi^A+i U^1_{A'}\phi^{A'})\,.
\label{Wscal}
\eea
The terms in the last line do not contribute to the WZ term, as they contain two space-time derivatives acting on the same scalar. Substituting the remaining terms into (\ref{G}) and computing the integral over the Grassmann variables, we find
\bea
\Gamma_{\rm WZ}&=&\int d^4x dU dV\,H^{(4)}(V^1_D\varphi^D-i U^1_{D'} \phi^{D'})
\partial_{\alpha\dot\alpha}\varphi^A
\partial_{\beta\dot\beta}\varphi^B
\partial^{\beta\dot\alpha}\phi^{A'}
\partial^{\alpha\dot\beta}\phi^{B'}
\nn\\&&\times
(V^2_A+iV^3_A)(V^2_B-iV^3_B)(U^2_{A'}-iU^3_{A'})(U^2_{B'}+iU^3_{B'})
\eea
With $\partial_{\alpha\dot\alpha}=\sigma^m_{\alpha\dot\alpha}\partial_m$, we apply the trace formula for the sigma-matrices (\ref{tr-sigma}) and single out only the term with the antisymmetric $\varepsilon$-tensor,
\be
\Gamma_{\rm WZ}=-8i\varepsilon^{mnpq}\int d^4x dU dV\,H^{(4)}(V^1_D\varphi^D-i U^1_{D'} \phi^{D'})
\partial_m\varphi^A
\partial_n\varphi^B
\partial_p\phi^{A'}
\partial_q\phi^{B'}
V^2_A V^3_B U^2_{A'} U^3_{B'}\,.
\label{4.15}
\ee
Substituting the series decomposition (\ref{ser}) into (\ref{4.15}) and computing the integral over the $U$-harmonics by the rule (\ref{hint2}), we obtain
\bea
\label{4.16}
\Gamma_{\rm WZ}&=&-16c\,\varepsilon^{mnpq}\varepsilon_{{A'}{B'}{C'}}
\int d^4x dV\,
\sum_{n=0}^\infty (n+1)(n+2)(-1)^n\frac{(\phi^{D'}\phi^{D'})^n}{(V^1_D\varphi^D)^{2n+5}}
\\&&\times
\phi^{A'}
\partial_p\phi^{B'}
\partial_q\phi^{C'}
\partial_m\varphi^A
\partial_n\varphi^B
V^2_A V^3_B
\nn\\
&&\hspace{-30pt}=
-32c\,\varepsilon^{mnpq}\varepsilon_{{A'}{B'}{C'}}
\int d^4x dV\,
\frac{V^1_C\varphi^C}{[\phi^{D'}\phi^{D'}+(V^1_D\varphi^D)^2]^3}
\phi^{A'}
\partial_p\phi^{B'}
\partial_q\phi^{C'}
\partial_m\varphi^A
\partial_n\varphi^B
V^2_A V^3_B
\,.\nn
\eea
(As in (\ref{4.9}), the series allowed explicit resummation.)
Next, we expand the integrand in a series over $V^1_D\varphi^D$
and perform the integration over the $V$-harmonics in a similar way,
\bea
\Gamma_{\rm WZ}&=&
-8c\,\varepsilon^{mnpq}
\int d^4x\,
\frac{1}{(\phi^{D'}\phi^{D'})^3}
\sum_{n=0}^\infty \frac{(-1)^n(n+2)(n+1)}{2n+3}
\left(
\frac{\varphi^D\varphi^D}{\phi^{D'}\phi^{D'}}
\right)^n
\nn\\&&\times
(\varepsilon_{{A'}{B'}{C'}}
\phi^{A'}
\partial_p\phi^{B'}
\partial_q\phi^{C'})
(\varepsilon_{ABC}
\varphi^A
\partial_m\varphi^B
\partial_n\varphi^C)\,.
\eea
The series can be summed up, and we obtain the following result
\be
\label{4.17}
\Gamma_{\rm WZ}=
-2c\,\varepsilon^{mnpq}
\int d^4x\,\frac{f(z)}{(\phi^{D'}\phi^{D'})^3}
(\varepsilon_{{A'}{B'}{C'}}
\phi^{A'}
\partial_p\phi^{B'}
\partial_q\phi^{C'})
(\varepsilon_{ABC}
\varphi^A
\partial_m\varphi^B
\partial_n\varphi^C)\,,
\ee
where
\be
f(z)=\frac{z^2-1}{z^2(z^2+1)^2}+\frac{\arctan z}{z^3}\,,\qquad
z^2=\frac{\varphi^A\varphi^A}{\phi^{A'} \phi^{A'}}\,.
\ee

Let us now introduce the normalized scalars,
\be
Y^A=\frac{\varphi^A}{\sqrt{\varphi^B\varphi^B+\phi^{B'}\phi^{B'}}}\,,\qquad
Y^{A'}=\frac{\phi^{A'}}{\sqrt{\varphi^B\varphi^B+\phi^{B'}\phi^{B'}}}\,,
\ee
which lie on the unit five-sphere, $Y^A Y^A+Y^{A'} Y^{A'}=1$.  In terms of these scalars, the action (\ref{4.17}) reads
\be
\Gamma_{\rm WZ}
=-2c\,\varepsilon^{mnpq}
\int d^4x\,g(y)
(\varepsilon_{ABC} Y^A
\partial_p Y^B
\partial_q Y^C)
(\varepsilon_{{A'}{B'}{C'}} Y^{A'}
\partial_m Y^{B'}
\partial_n Y^{C'})\,,
\label{4.18}
\ee
where
\be
g(y)=\frac{y^4-1}{y^2}+\frac{(y^2+1)^3}{y^3}\arctan y\,,\qquad
y^2=\frac{Y^A Y^A}{Y^{A'} Y^{A'}}\,.
\ee
Comparing (\ref{4.18}) with (\ref{1}), we see that we have perfect agreement provided
\be
c=\frac1{32\pi^2}\,.
\ee
The $F^4/X^4$ term (\ref{F4X4}) then also has the coefficient as given in \cite{BS}. According to the analysis presented in \cite{BS} (see also references quoted there), this is the minimal value of the constant $c$ allowed by the topological quantization condition. It corresponds to the case when the SYM gauge group SU(2) is broken to U(1).

\subsection{SU($N$) gauge group}
The effective action (\ref{G}) with the function $H(W)$ given in (\ref{H}) can be easily generalized to describe the $\cN=4$ SYM effective action on the Coulomb branch in the case when the gauge group SU($N$) is spontaneously broken to its maximal abelian subgroup $[{\rm U}(1)]^{N-1}$. The superfield $W$ in this case is a traceless diagonal $N\times N$ matrix in the Cartan subalgebra of $su(N)$,
\be
W={\rm diag}(W^1,W^1,\ldots,W^N)\,,\qquad
\sum_{i=1}^N W^i=0\,,
\ee
with all eigenvalues being distinct: $W^i\neq W^j$ if $i\neq j$. The effective action in this case reads
\be
\Gamma=\frac1{32\pi^2}\int d\zeta du dv\sum_{i<j}^N\ln
\frac{|W^i-W^j|}{\Lambda}\,.
\ee
For the case of the gauge group SU(2) spontaneously broken down to U(1)
this sum reduces to (\ref{G}) with (\ref{H}). For non-unitary gauge
groups, the effective action can be written in a similar way with the
summation over the positive roots of the gauge algebra (see, e.g.,
\cite{BuIv,BP2,BBK,BuSams1,BuSams2} for similar generalizations in the $\cN=2$ superspace).

\section{Conclusions}

In this paper, we developed a formulation of the $\cN=4$ SYM
effective action in a novel $\cN=4$\, $d=4$ bi-harmonic
superspace. The coordinates of this superspace involve two copies
of the standard SU(2) harmonic variables, $u^\pm_i$ and $v^\pm_a$,
which allowed us to make manifest the
SU(2)$\times$SU(2)$\sim$SO(3)$\times$SO(3) subgroup of the full
SU(4)$\sim$SO(6) R-symmetry group of the $\cN=4$ SYM theory.

The idea of introducing this superspace was inspired by our
previous work \cite{BS}, where we showed that the effective
Lagrangian in the $\cN=4$ SYM theory written in the
four-dimensional form can be made manifestly invariant under one
of the following three subgroups of the full SO(6) R-symmetry:
SO(5), SO(4)$\times$SO(2) or SO(3)$\times$SO(3). In \cite{BS}, we
explored two superspace formulations of the $\cN=4$ SYM effective
action which correspond to the SO(5) and SO(4)$\times$SO(2)
subgroups, whereas in the present paper we constructed a novel
superspace which makes the SO(3)$\times$SO(3) subgroup manifest.
We therefore demonstrated that for each of the maximal
non-anomalous subgroups of the SU(4) R-symmetry there exists the
corresponding superspace description of the $\cN=4$ SYM effective
action.

Representing the SU(4) index $I$ by a pair of SU(2) indices
$(i,a)$, we found that the antisymmetric $\cN=4$ superfield
strength $W^{IJ}$ is equivalently described by two
\emph{symmetric} SU(2) tensors, $W^{(ij)}$ and $W^{(ab)}$.  We
found constraints for these superfields which restrict their field
content to that of the on-shell $\cN=4$ supergauge multiplet and
which are equivalent to the standard constraints for the
superfield $W^{IJ}$. At this stage, the constraints still mix all
the superfield components, so that one needs both $W^{(ij)}$ and
$W^{(ab)}$ to describe the $\cN=4$ SYM multiplet.

As the next step, we contracted $W^{(ij)}$ and $W^{(ab)}$ with the
harmonic variables and obtained six $\cN=4$ superfields $W$, $W'$,
$W^{(\pm2,0)}$, $W^{(0,\pm2)}$ with definite charges under the
U(1)$\times$U(1) subgroup of SU(2)$\times$SU(2). We then found
that we could \emph{decouple} the corresponding constraints, so
that any one of these six bi-harmonic superfields can be used to
describe the $\cN=4$ SYM multiplet. The advantage of using these
superfields instead of $W^{IJ}$ is that part of the constraints
can be solved by simply passing to the corresponding analytic
subspaces. We explicitly solved the remaining constraints for the
superfields $W$ and $W^{(2,0)}$, and gave the corresponding
component field decompositions. We showed that all the auxiliary
field components vanish, whereas the physical components
correspond to the on-shell $\cN=4$ gauge multiplet.

We argued that the chargeless harmonic superfield $W$ is
particularly convenient for describing the low-energy effective
$\cN=4$ SYM action. We explicitly demonstrated this by presenting
the leading four-derivative part of the effective action in the
bi-harmonic superspace. Scale invariance required this part to be
just $\ln W$ integrated over the corresponding analytic subspace.
The similarity of this result with the corresponding expression in
a totally different $\cN=4$ USp(4) harmonic superspace, see
\cite{BS}, is quite striking. Even more striking is the way the
algebra worked out in our proof that the expected $F^4/X^4$ and WZ
terms are indeed contained in this superfield expression.

It is the main (and somewhat unexpected) result of the present
paper that the $\cN=4$ SYM low-energy effective action becomes
extremely simple when considered in the $\cN=4$ bi-harmonics
superspace.

It still remains to be understood whether and how this effective
action could be described in the $\cN=4$ SU(4) harmonic superspace
\cite{Npq,HH,HW1,HW2} or in the $\cN=3$ SU(3) harmonic superspace
\cite{Galperin:1985uw,Galperin:1984bu}. In \cite{BS}, we pointed
out that these are \emph{not} particularly convenient superspaces
because SU(4) and SU(3) have `mild' anomalies that, via the
properties of the WZ term, prevent these symmetries to be
manifestly realized. (They are still symmetries of the effective
action, but must transform the Lagrangian into a total
divergence.) However, the corresponding descriptions could still
exist.

An important application of our results would be to try to
construct higher-derivative parts of the effective action in the
available $\cN=4$ superspaces introduced in \cite{BLS,BS} and in
the present paper. It would also be interesting to see if the
on-shell restrictions could be relaxed in these superspaces.
Then one could try to rederive our results through explicit perturbative quantum calculations in the off-shell harmonic superspaces (cf.~\cite{BK}).
We leave these problems for future studies.

\vspace{30pt}
\noindent
{\bf Acknowledgments}\\[3mm]
The research of D.B. was supported by the Department of Energy Grant No.\ DE-FG02-97ER41029. The work of I.S.\ was supported by the Marie Curie research fellowship No.\ 236231, ``QuantumSupersymmetry''; by RFBR grants Nr.\ 09-02-00078 and 11-02-90445; by grant for LRSS, project No.\ 3558.2010.2.

\appendix

\section{Covariant derivatives in analytic coordinates}

In this appendix, we give the explicit form of the covariant spinor and harmonic derivatives in the analytic subspaces defined in (\ref{table}).

\subsection{Spinor and harmonic derivatives in the subspace $A_1$}

Covariant spinor derivatives:
\bea
&&D^{\2}_\alpha=-\frac\partial{\partial\theta^{\2\alpha}}
 \,,\qquad
D^{\1}_\alpha=\frac\partial{\partial\theta^{\1\alpha}}
 +2i\bar\theta^{\1\dot\alpha}\partial_{\alpha\dot\alpha}\,,\nn\\
&&D^\3_\alpha=-\frac\partial{\partial\theta^{\3\alpha}}
 \,,\qquad
D^{\4}_\alpha=\frac\partial{\partial\theta^{\4\alpha}}
+2i\bar\theta^{\4\dot\alpha}\partial_{\alpha\dot\alpha} \,,\nn\\
&&\bar D^{\1}_{\dot\alpha}=-\frac\partial{\partial\bar\theta^{\1\dot\alpha}}
\,,\qquad
\bar D^{\2}_{\dot\alpha}=\frac\partial{\partial\bar\theta^{\2\dot\alpha}}
-2i\theta^{\2\alpha}\partial_{\alpha\dot\alpha}\,,\nn\\
&&\bar D^{\4}_{\dot\alpha}=-\frac\partial{\partial\bar\theta^{\4\dot\alpha}}\,,\qquad
\bar D^{\3}_{\dot\alpha}=\frac\partial{\partial\bar\theta^{\3\dot\alpha}}
-2i\theta^{\3\alpha}\partial_{\alpha\dot\alpha}
\,.
\eea
Covariant harmonic derivatives:
\bea
D^{(2,0)}_{A_1}&=&D^{(2,0)}
+2i\theta^{\2\alpha}\bar\theta^{\4\dot\alpha}\partial_{\alpha\dot\alpha}
+2i\theta^{\1\alpha}\bar\theta^{\3\dot\alpha}\partial_{\alpha\dot\alpha}
+\theta^{\1}_\alpha\frac\partial{\partial\theta^{\3}_\alpha}
+\theta^{\2}_\alpha\frac\partial{\partial\theta^{\4}_\alpha}
+\bar\theta^{\4}_{\dot\alpha}\frac\partial{\partial\bar\theta^{\2}_{\dot\alpha}}
+\bar\theta^{\3}_{\dot\alpha}\frac\partial{\partial\bar\theta^{\1}_{\dot\alpha}}
\,,
\nn\\
D^{(-2,0)}_{A_1}&=&D^{(-2,0)}
+2i\theta^{\4\alpha}\bar\theta^{\2\dot\alpha}\partial_{\alpha\dot\alpha}
+2i\theta^{\3\alpha}\bar\theta^{\1\dot\alpha}\partial_{\alpha\dot\alpha}
+\theta^{\3}_\alpha\frac\partial{\partial\theta^{\1}_\alpha}
+\theta^{\4}_\alpha\frac\partial{\partial\theta^{\2}_\alpha}
+\bar\theta^{\2}_{\dot\alpha}\frac\partial{\partial\bar\theta^{\4}_{\dot\alpha}}
+\bar\theta^{\2}_{\dot\alpha}\frac\partial{\partial\bar\theta^{\3}_{\dot\alpha}}\,,
\nn\\
D^{(0,2)}_{A_1}&=&D^{(0,2)}+2i\theta^{\1\alpha} \bar\theta^{\2\dot\alpha}
\partial_{\alpha\dot\alpha}
+2i\theta^{\3\alpha}\bar\theta^{\4\dot\alpha}\partial_{\alpha\dot\alpha}
+\theta^{\1}_\alpha\frac\partial{\partial\theta^{\2}_\alpha}
+\theta^{\3}_\alpha\frac\partial{\partial\theta^{\4}_\alpha}
+\bar\theta^{\4}_{\dot\alpha}\frac\partial{\partial\bar\theta^{\3}_{\dot\alpha}}
+\bar\theta^{\2}_{\dot\alpha}\frac\partial{\partial\bar\theta^{\2}_{\dot\alpha}}\,,
\nn\\
D^{(0,-2)}_{A_1}&=&D^{(0,-2)}+2i\theta^{\4\alpha} \bar\theta^{\3\dot\alpha}
\partial_{\alpha\dot\alpha}
+2i\theta^{\2\alpha}\bar\theta^{\1\dot\alpha}\partial_{\alpha\dot\alpha}
+\theta^{\2}_\alpha\frac\partial{\partial\theta^{\1}_\alpha}
+\theta^{\4}_\alpha\frac\partial{\partial\theta^{\3}_\alpha}
+\bar\theta^{\3}_{\dot\alpha}\frac\partial{\partial\bar\theta^{\4}_{\dot\alpha}}
+\bar\theta^{\1}_{\dot\alpha}\frac\partial{\partial\bar\theta^{\2}_{\dot\alpha}}\,.
\nn\\
\label{HA1}
\eea
In the above expressions, $\partial_{\alpha\dot\alpha}=\sigma^m_{\alpha\dot\alpha}\frac\partial{\partial
x^m_{A_1}}$.

\subsection{Spinor and harmonic derivatives in the subspace $A_2$}
Covariant spinor derivatives:
\bea
&&\bar D^{\1}_{\dot\alpha}=-\frac\partial{\partial\bar\theta^{\1\dot\alpha}}
\,,\qquad
\bar D^{\3}_{\dot\alpha}=+\frac\partial{\partial\bar\theta^{\3\dot\alpha}}
-2i\theta^{\3\alpha}\partial_{\alpha\dot\alpha}
\,,\nn\\
&&\bar D^{\2}_{\dot\alpha}=+\frac\partial{\partial\bar\theta^{\2\dot\alpha}}\,,\qquad
\bar D^{\4}_{\dot\alpha}=-\frac\partial{\partial\bar\theta^{\4\dot\alpha}}
-2i\theta^{\4\alpha}\partial_{\alpha\dot\alpha}\,,\nn\\
&&D^{\3}_\alpha=-\frac\partial{\partial\theta^{\3\alpha}}\,,\qquad
D^{\1}_\alpha=+\frac\partial{\partial\theta^{\1\alpha}}
 +2i\bar\theta^{\1\dot\alpha}\partial_{\alpha\dot\alpha}\,,\nn\\
&&D^{\4}_\alpha=+\frac\partial{\partial\theta^{\4\alpha}}
\,,\qquad
D^{\2}_\alpha=-\frac\partial{\partial\theta^{\2\alpha}}
 +2i\bar\theta^{\2\dot\alpha}\partial_{\alpha\dot\alpha}
\,.
\eea
Covariant harmonic derivatives:
\bea
\label{A4}
D^{(2,0)}_{A_2}&=&D^{(2,0)}
-2i\theta^{\2\alpha}\bar\theta^{\4\dot\alpha}\partial_{\alpha\dot\alpha}
+2i\theta^{\1\alpha}\bar\theta^{\3\dot\alpha}\partial_{\alpha\dot\alpha}
+\theta^{\1}_\alpha\frac\partial{\partial\theta^{\3}_\alpha}
+\theta^{\2}_\alpha\frac\partial{\partial\theta^{\4}_\alpha}
+\bar\theta^{\4}_{\dot\alpha}\frac\partial{\partial\bar\theta^{\2}_{\dot\alpha}}
+\bar\theta^{\3}_{\dot\alpha}\frac\partial{\partial\bar\theta^{\1}_{\dot\alpha}}
\,,
\nn\\
D^{(-2,0)}_{A_2}&=&D^{(-2,0)}
-2i\theta^{\4\alpha}\bar\theta^{\2\dot\alpha}\partial_{\alpha\dot\alpha}
+2i\theta^{\3\alpha}\bar\theta^{\1\dot\alpha}\partial_{\alpha\dot\alpha}
+\theta^{\3}_\alpha\frac\partial{\partial\theta^{\1}_\alpha}
+\theta^{\4}_\alpha\frac\partial{\partial\theta^{\2}_\alpha}
+\bar\theta^{\2}_{\dot\alpha}\frac\partial{\partial\bar\theta^{\4}_{\dot\alpha}}
+\bar\theta^{\1}_{\dot\alpha}\frac\partial{\partial\bar\theta^{\3}_{\dot\alpha}}\,,
\nn\\
D^{(0,2)}_{A_2}&=&D^{(0,2)}
+\theta^{\1}_\alpha\frac\partial{\partial\theta^{\2}_\alpha}
+\theta^{\3}_\alpha\frac\partial{\partial\theta^{\4}_\alpha}
+\bar\theta^{\4}_{\dot\alpha}\frac\partial{\partial\bar\theta^{\3}_{\dot\alpha}}
+\bar\theta^{\2}_{\dot\alpha}\frac\partial{\partial\bar\theta^{\1}_{\dot\alpha}}\,,
\nn\\
D^{(0,-2)}_{A_2}&=&D^{(0,-2)}
+\theta^{\2}_\alpha\frac\partial{\partial\theta^{\1}_\alpha}
+\theta^{\4}_\alpha\frac\partial{\partial\theta^{\3}_\alpha}
+\bar\theta^{\3}_{\dot\alpha}\frac\partial{\partial\bar\theta^{\4}_{\dot\alpha}}
+\bar\theta^{\1}_{\dot\alpha}\frac\partial{\partial\bar\theta^{\2}_{\dot\alpha}}\,.
\eea
In the above expressions, $\partial_{\alpha\dot\alpha}=\sigma^m_{\alpha\dot\alpha}\frac\partial{\partial
x^m_{A_2}}$.

\section{SO(3) harmonic integrals}
The SO(3) harmonic variables $U$ and $V$ are nothing but the usual SO(3)
matrices with the properties (\ref{SO3harm}).
The relations (\ref{B2}) among these harmonics and the SU(2) ones,
together with the SU(2) harmonic integration rules (\ref{u-int}), yield
\bea
&&\int dU\,1=1\,,\qquad
\int dU(\mbox{non-singlet SO(3) irreducible representation})=0\,,
\label{SO3-int}\nn\\&&
\int dV\,1=1\,,\qquad
\int dV(\mbox{non-singlet SO(3) irreducible representation})=0\,. \hspace{20pt}
\eea
There are two obvious consequences of these SO(3) harmonic integration rules,
\be
\int dV\, V^1_A V^1_B=\frac13\delta_{AB}\,,\qquad
\int dV\, V^1_A V^2_B V^3_C=\frac1{3!}\varepsilon_{ABC}\,,
\ee
where $\varepsilon_{ABC}$ is the totally antisymmetric SO(3) tensor.
After a bit of combinatorics, we obtain the following generalization of these two
harmonic integrals
\bea
\int dV\, V^1_{A_1}\ldots V^1_{A_k}&=&\left\{
\begin{array}{ll}
\dfrac1{k+1}\delta_{(A_1 A_2} \ldots
\delta_{A_{k-1}A_k )}\,, & k=2n\\[10pt]
0&k=2n+1\,,
\end{array}
\right.
\label{hint1}
\\
\int dV\, V^1_{A_1}\ldots V^1_{A_{k}}V^1_A V^2_B V^3_C
&=&\left\{
\begin{array}{ll}
\dfrac{\delta_{(A_1 A_2}\ldots
\delta_{A_{k-1}A_{k}}\varepsilon_{A)BC}}{2(k+3)}\,,& k=2n\\
0& k=2n+1\,.
\end{array}
\right.
\label{hint2}
\eea
The same identities hold also for the integrals with the $U$-harmonics.




\begin{thebibliography}{99}

\bibitem{BS} D.~V.~Belyaev and I.~B.~Samsonov,
  {\it Wess-Zumino term in the N=4 SYM effective action
  revisited}, JHEP {\bf 1104} (2011) 112, {\tt arXiv:1103.5070
  [hep-th]}.

\bibitem{DS}
  M.~Dine and N.~Seiberg,
  {\it Comments on higher derivative operators in some SUSY field
  theories},
  Phys.\ Lett.\  B {\bf 409} (1997) 239, {\tt arXiv:hep-th/9705057}.

\bibitem{S16}
  N.~Seiberg,
  {\it Notes on theories with 16 supercharges},
  Nucl.\ Phys.\ Proc.\ Suppl.\  {\bf 67} (1998) 158,
  {\tt arXiv:hep-th/9705117}.

\bibitem{TZ}
  A.~A.~Tseytlin and K.~Zarembo,
  {\it Magnetic interactions of D-branes and Wess-Zumino terms in super
  Yang-Mills effective actions},
  Phys.\ Lett.\  B {\bf 474} (2000) 95, {\tt arXiv:hep-th/9911246}.

\bibitem{Intriligator:2000eq}
  K.~A.~Intriligator,
  {\it Anomaly matching and a Hopf-Wess-Zumino term in 6d, N=(2,0) field
  theories},
  Nucl.\ Phys.\  B {\bf 581} (2000) 257, {\tt arXiv:hep-th/0001205}.

\bibitem{BLS}
  I.~L.~Buchbinder, O.~Lechtenfeld and I.~B.~Samsonov,
  {\it N=4 superparticle and super Yang-Mills theory in USp(4) harmonic
  superspace},
  Nucl.\ Phys.\  B {\bf 802} (2008) 208, {\tt arXiv:0804.3063 [hep-th]}.

\bibitem{Claus}
  P.~Claus, R.~Kallosh, J.~Kumar, P.~K.~Townsend and A.~Van Proeyen,
  {\it Conformal theory of M2, D3, M5 and D1+D5 branes},
  JHEP {\bf 9806} (1998) 004, {\tt arXiv:hep-th/9801206}.

\bibitem{Braaten}
  E.~Braaten, T.~L.~Curtright and C.~K.~Zachos,
  {\it Torsion and geometrostasis in nonlinear sigma models},
  Nucl.\ Phys.\  B {\bf 260} (1985) 630.

\bibitem{GIKOS}
  A.~Galperin, E.~Ivanov, S.~Kalitzin, V.~Ogievetsky and E.~Sokatchev,
  {\it Unconstrained N=2 matter, Yang-Mills and supergravity theories in harmonic
  superspace},
  Class.\ Quant.\ Grav.\  {\bf 1} (1984) 469.

\bibitem{book}
  A.~S.~Galperin, E.~A.~Ivanov, V.~I.~Ogievetsky and E.~S.~Sokatchev,
  {\it Harmonic Superspace},
 Cambridge, UK: Univ. Pr. (2001) 306 p.

\bibitem{BuIv}
  I.~L.~Buchbinder and E.~A.~Ivanov,
  {\it Complete N=4 structure of low-energy effective action in N=4 super
  Yang-Mills theories},
  Phys.\ Lett.\  B {\bf 524} (2002) 208, {\tt arXiv:hep-th/0111062}.

\bibitem{BIP}
  I.~L.~Buchbinder, E.~A.~Ivanov and A.~Y.~Petrov,
  {\it Complete low-energy effective action in N=4 SYM: A direct N=2 supergraph
  calculation},
  Nucl.\ Phys.\  B {\bf 653} (2003) 64,
  {\tt arXiv:hep-th/0210241}.

\bibitem{BBP}
  A.~T.~Banin, I.~L.~Buchbinder and N.~G.~Pletnev,
  {\it One loop effective action for N=4 SYM theory in the hypermultiplet sector:
  Leading low-energy approximation and beyond},
  Phys.\ Rev.\  D {\bf 68} (2003) 065024,
  {\tt arXiv:hep-th/0304046}.

\bibitem{BP1}
  I.~L.~Buchbinder and N.~G.~Pletnev,
  {\it Construction of one-loop N=4 SYM effective action on the mixed branch in
  the harmonic superspace approach},
  JHEP {\bf 0509} (2005) 073,
  {\tt arXiv:hep-th/0504216}.

\bibitem{BP2}
  I.~L.~Buchbinder and N.~G.~Pletnev,
  {\it Hypermultiplet dependence of one-loop effective action in the N=2
  superconformal theories},
  JHEP {\bf 0704} (2007) 096,
  {\tt arXiv:hep-th/0611145}.

\bibitem{IS1}
  E.~Ivanov and A.~Sutulin,
  {\it Sigma models in (4,4) harmonic superspace},
  Nucl.\ Phys.\  B {\bf 432} (1994) 246
  [Erratum-ibid.\  B {\bf 483} (1997) 531],
  {\tt arXiv:hep-th/9404098}.

\bibitem{Ivanov:1995jb}
  E.~A.~Ivanov,
  {\it On the harmonic superspace geometry of (4,4) supersymmetric sigma models
  with torsion},
  Phys.\ Rev.\  D {\bf 53} (1996) 2201,
  {\tt arXiv:hep-th/9502073}.

\bibitem{Ivanov:1995yp}
  E.~A.~Ivanov,
  {\it Off-shell (4,4) supersymmetric sigma models with torsion as gauge theories
  in harmonic superspace},
  Phys.\ Lett.\  B {\bf 356} (1995) 239,
  {\tt arXiv:hep-th/9504070}.

\bibitem{BI}
  S.~Bellucci and E.~Ivanov,
  {\it N=(4,4), 2-D supergravity in SU(2)$\times$SU(2) harmonic superspace},
  Nucl.\ Phys.\  B {\bf 587} (2000) 445,
  {\tt arXiv:hep-th/0003154}.

\bibitem{IS2}
  E.~Ivanov and A.~Sutulin,
  {\it Diversity of off-shell twisted (4,4) multiplets in SU(2)$\times$SU(2) harmonic
  superspace},
  Phys.\ Rev.\  D {\bf 70} (2004) 045022, {\tt arXiv:hep-th/0403130}.

\bibitem{BIS}
  S.~Bellucci, E.~Ivanov and A.~Sutulin,
  {\it N=8 mechanics in SU(2)$\times$SU(2) harmonic superspace},
  Nucl.\ Phys.\  B {\bf 722} (2005) 297
  [Erratum-ibid.\  B {\bf 747} (2006) 464], {\tt arXiv:hep-th/0504185}.

\bibitem{IN}
  E.~Ivanov and J.~Niederle,
  {\it Bi-harmonic superspace for N=4 mechanics},
  Phys.\ Rev.\  D {\bf 80} (2009) 065027, {\tt arXiv:0905.3770 [hep-th]}.


\bibitem{S1}
  M.~F.~Sohnius,
  {\it Bianchi identities for supersymmetric gauge theories},
  Nucl.\ Phys.\  B {\bf 136} (1978) 461.

\bibitem{S2}
  M.~F.~Sohnius,
  {\it Supersymmetry and central charges},
  Nucl.\ Phys.\  B {\bf 138} (1978) 109.

\bibitem{BBK}
  E.~I.~Buchbinder, I.~L.~Buchbinder and S.~M.~Kuzenko,
  {\it Nonholomorphic effective potential in N=4 SU(n) SYM},
  Phys.\ Lett.\  B {\bf 446} (1999) 216,
  {\tt arXiv:hep-th/9810239}.

\bibitem{BuSams1}
  I.~L.~Buchbinder and I.~B.~Samsonov,
  {\it On holomorphic effective actions of hypermultiplets coupled to external
  gauge superfields},
  Mod.\ Phys.\ Lett.\  A {\bf 14} (1999) 2537,
  {\tt arXiv:hep-th/9909183}.

\bibitem{BuSams2}
  I.~L.~Buchbinder and I.~B.~Samsonov,
  {\it The holomorphic effective action in N=2 D = 4 supergauge theories with
  various gauge groups},
  Theor.\ Math.\ Phys.\  {\bf 122} (2000) 371
  [Teor.\ Mat.\ Fiz.\  {\bf 122} (2000) 444].


\bibitem{Npq}
  G.~G.~Hartwell and P.~S.~Howe,
  {\it (N,p,q) harmonic superspace},
  Int.\ J.\ Mod.\ Phys.\  A {\bf 10} (1995) 3901,
  {\tt arXiv:hep-th/9412147}.

\bibitem{HH}
  P.~S.~Howe and G.~G.~Hartwell,
  {\it A Superspace survey},
  Class.\ Quant.\ Grav.\  {\bf 12} (1995) 1823.

\bibitem{HW1}
  P.~S.~Howe and P.~C.~West,
  {\it Nonperturbative Green's functions in theories with extended superconformal
  symmetry},
  Int.\ J.\ Mod.\ Phys.\  A {\bf 14} (1999) 2659,
  {\tt arXiv:hep-th/9509140}.

\bibitem{HW2}
  P.~S.~Howe and P.~C.~West,
  {\it Superconformal invariants and extended supersymmetry},
  Phys.\ Lett.\  B {\bf 400} (1997) 307,
  {\tt arXiv:hep-th/9611075}.

\bibitem{Galperin:1985uw}
  A.~Galperin, E.~Ivanov, S.~Kalitsyn, V.~Ogievetsky, E.~Sokatchev,
  {\it N = 3 supersymmetric gauge theory},
  Phys.\ Lett.\  {\bf B151} (1985) 215.

\bibitem{Galperin:1984bu}
  A.~Galperin, E.~Ivanov, S.~Kalitsyn, V.~Ogievetsky, E.~Sokatchev,
  {\it Unconstrained off-shell N=3 supersymmetric Yang-Mills theory},
  Class.\ Quant.\ Grav.\  {\bf 2} (1985) 155.

\bibitem{BK}
  I.~L.~Buchbinder and S.~M.~Kuzenko,
  {\it Comments on the background field method in harmonic superspace:
  Nonholomorphic corrections in N=4 SYM},
  Mod.\ Phys.\ Lett.\  A {\bf 13} (1998) 1623,
  {\tt arXiv:hep-th/9804168}.


\end{thebibliography}
\end{document}